\documentclass[aps,showpacs,floatfix,twocolumn]{revtex4}
\usepackage{graphicx}
\usepackage{epsfig}
\usepackage{amsmath}
\usepackage{graphics}
\usepackage{latexsym}
\usepackage{adjustbox}
\begin{document}
\title{Linear Dimensions of Adsorbed Semiflexible Polymers:\\ What can be 
learned  about their persistence length?}
\author{Andrey Milchev$^{1,2}$}
\author{Kurt Binder$^2$}
\affiliation{$^1$Institute for Physical Chemistry, Bulgarian Academia of 
Sciences, 1113, Sofia, Bulgaria}
\affiliation{$^2$Institute of Physics, Johannes Gutenberg University Mainz,
Staudingerweg 7, 55128 Mainz, Germany}

\begin{abstract}
Conformations of partially or fully adsorbed semiflexible polymer chains are 
studied varying both contour length $L$, chain stiffness, $\kappa$, and the 
strength of the adsorption potential over a wide range. Molecular Dynamics 
simulations show that partially adsorbed chains (with ``tails'', surface 
attached ``trains'' and ``loops'') are not described by the Kratky-Porod 
wormlike chain model. The crossover of the persistence length from its 
three-dimensional value $(\ell_p)$ to the enhanced value in two dimensions 
$(2\ell_p)$ is analyzed, and excluded volume effects are identified for $L \gg  
\ell_p$. Consequences for the interpretation of experiments are suggested. We 
verify the prediction that the adsorption threshold scales as $\ell_p^{-1/3}$.
\end{abstract}
\maketitle

\begin{figure}[b]
\begin{adjustbox}{center}
\hspace{-1.0cm}
 \includegraphics[scale=0.25, angle=0]{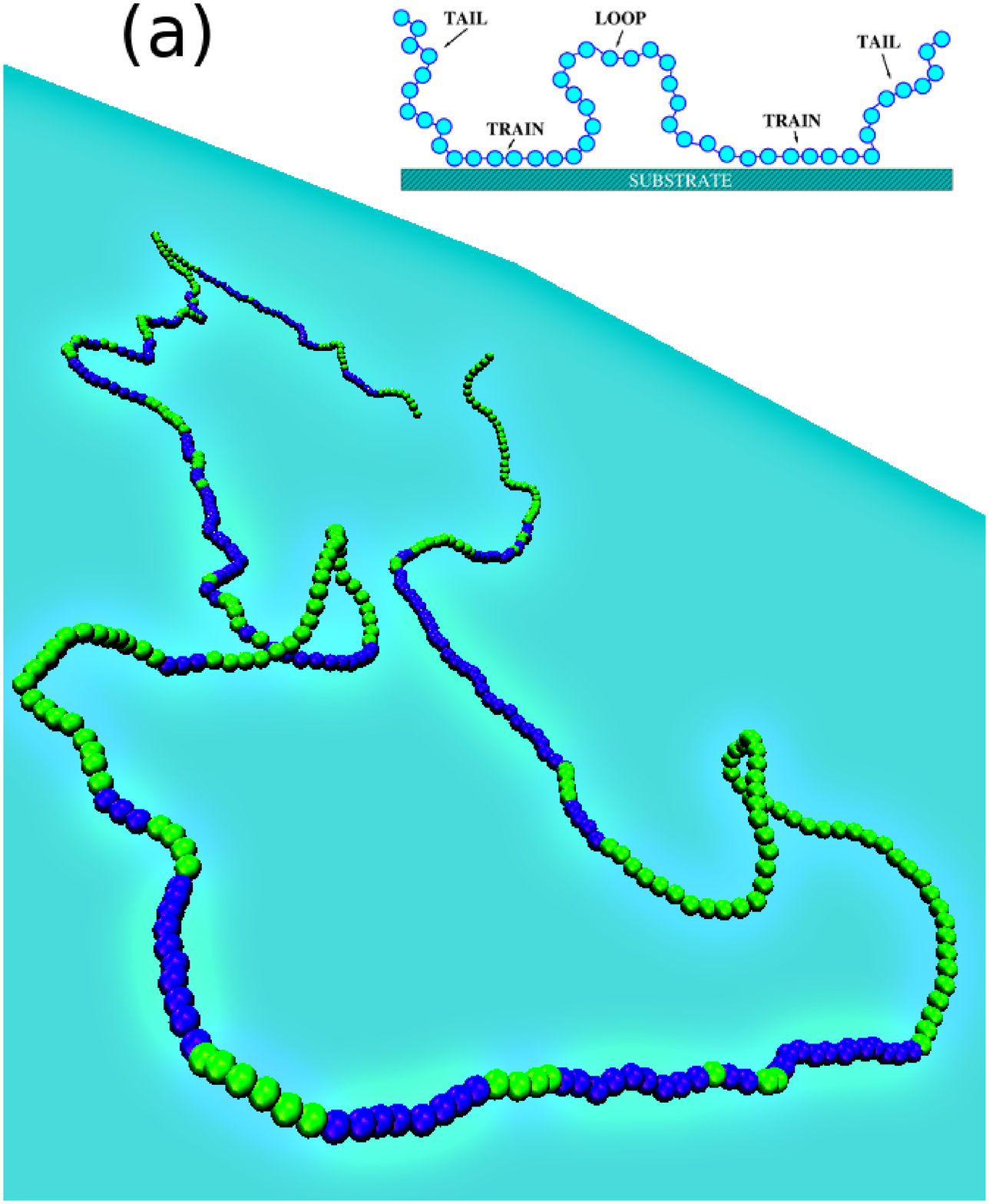}
\end{adjustbox}

 \vspace{0.7cm} 
\hspace{-1.0cm}
 \begin{adjustbox}{center}
 \includegraphics[scale=0.25, angle=0]{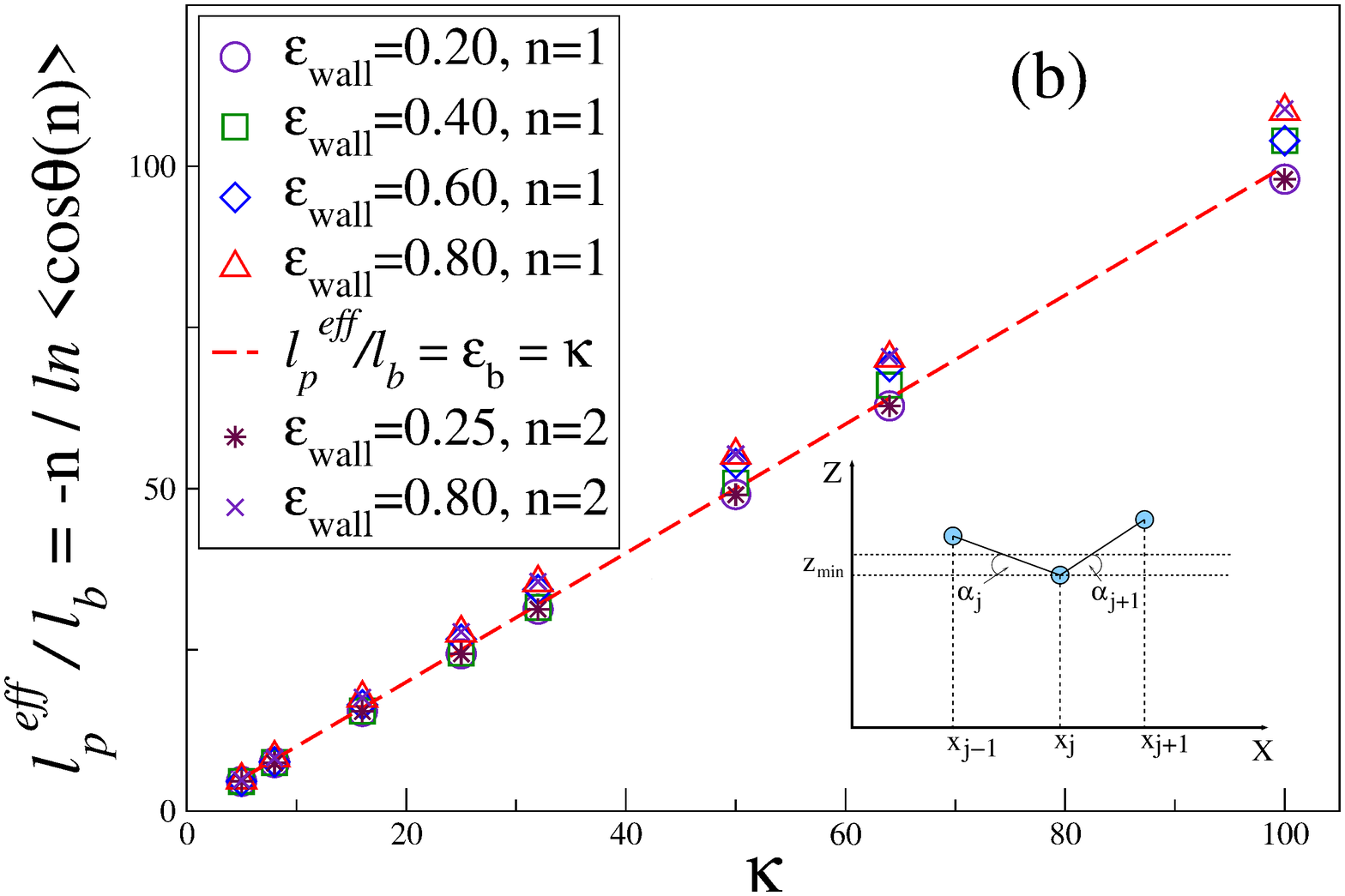}
\end{adjustbox}
\caption{(a) Snapshot of an adsorbed chain with $N=500$ for $\kappa=16, 
\epsilon_{wall}=0.65$. Loops and a tail are shown in green, trains are in 
darkblue. (b) Decay length $\ell_p^{eff} / \ell_b$ vs stiffness $\kappa$ for 
$N=250$ and several choices of $\epsilon_{wall}$. Here $n=1$ means an angle 
between nearest bonds,  $n=2$ stands for next-nearest bonds. Data for $n=1,\,2$ 
indicate that $\ell_p^{eff}$ increases rather gradually with $\kappa$ for 
adsorbed chains. The inset illustrates the geometry of the $x,z$-coordinates of 
two subsequent bonds where the $X$-axis is chosen such that the bond from 
$\vec{r}_{j-1}$ to $\vec{r}_j$ lies in the $X,Z$ plane. The angles $\alpha_j = 
\frac{\pi}{2}- \vartheta_j$ are the complements to the polar angles 
$\vartheta_j$  of the bonds with the $Z$-axis.}
\label{fig_snapshot}
\par\vspace{-0.7cm}\par
\end{figure}

{\em \bf Introduction} 
Adsorbed stiff macromolecules on substrates are of key interest to understand 
properties and function of various nanomaterials, and also play an important 
role in biological context 
\cite{Wiggins,Han,Ha,Moukhtar1,Moukhtar2,Radler,Rechendorff}. While adsorption 
of flexible polymers has been extensively studied 
\cite{Sinha,deGennes,Eisenriegler,Fleer,Klushin}, the adsorption transition of 
semiflexible polymers is much less understood 
\cite{Birshtein,Khokhlov,Kramarenko,Maggs,Semenov,Netz,Deng,Hsu1,Kampmann,
Baschnagel,Kierfeld}. For flexible polymers, the salient features of this 
transition are well captured \cite{Eisenriegler,Fleer,Klushin} by the simple 
selfavoiding walk lattice model of polymers \cite{Grosberg}.  However, extending 
the model to semiflexible polymers \cite{Birshtein,Hsu1} misses important 
degrees of freedom, namely, chain bending \cite{Odijk} by small bending angle 
$\theta$. Consequently, most work uses the Kratky-Porod (KP)\cite{KP} wormlike 
chain (WLC) model: in the continuum limit the chain is described by a curve 
$\vec{r}(s)$ in space, the only energy parameter $\kappa$ considered relates to 
the local curvature of the polymer. The Hamiltonian
\begin{equation} \label{H}
 \frac{\cal H}{k_BT} = \frac{\kappa}{2} \int_0^L ds \left( 
\frac{d^2\vec{r}(s)}{ds^2} \right)^2
\end{equation}
yields for the tangent - tangent correlation function an exponential decay 
with the distance $n$ between two bond vectors ($n = s-s'$) along the chain 
backbone,
\begin{equation} \label{lp}
 C(n) = \langle \cos\theta(n)\rangle = e^{-n/\ell_p} (d=3), \mbox{or}\; 
e^{-n/2\ell_p}\,\, (d=2),
\end{equation}
with $\ell_p = \kappa$ the persistence length. There are two problems: (i) 
while in $d=3$ dimensions excluded volume interactions between the effective 
monomer units of the polymer come into play only for extremely long chains when 
$\ell_p \gg 1$ (measuring lengths in units of the distance $\ell_b=1$ between 
the subsequent monomers along the chain) \cite{Hsu2}, in $d=2$ deviations from 
Eq.(\ref{lp}) start when $s-s'$ exceeds $2\ell_p$ distinctly, and a gradual 
crossover to a power-law decay, $\langle \cos \theta(s-s') \rangle \propto 
(s-s')^{-\beta}$ with $\beta = 2(1-\nu)=1/2$ \cite{Hsu3} begins. Strictly in 
$d=2$, chains cannot intersect, and for $L \gg \ell_p$ excluded volume matters. 
(ii) in fact, adsorbed chains exist to some extent ``in between'' the dimensions 
(remember the well-known \cite{Fleer} description in terms of trains, tails and 
loops, cf. Fig.\ref{fig_snapshot}a: tails and loops exists in $d=3$, trains 
reside (almost) in $d=2$). If the adsorption potential, $U(z)$, with $z$ being 
the distance from the (planar)  adsorbing substrate, is very strong, tails and 
loops will be essentially eliminated but in real systems the adsorption then 
must be expected to be irreversible\cite{Lee}. While single-stranded (ss)-DNA on 
graphite \cite{Rechendorff} and double-stranded (ds)-DNA on lipid membranes 
\cite{Radler} have been shown to equilibrate by diffusion in the adsorbed state, 
no diffusion is observed for more bulky polymers such as dendronized polymers 
(DP) \cite{Grebikova}. Adsorbed bottlebrush polymers \cite{Hsu4} or DPs are 
intriguing since $\ell_p$ for such polymers can be systematically varied by 
choosing different side chain lengths (for bottlebrushes \cite{Hsu5}), or 
different generations (for DPs  \cite{Grebikova,Messner,Duterte}). However, 
experiments reveal subtle effects of surface roughness \cite{Grebikova} and 
electrostatic interactions \cite{Grebikova} making thus the interpretation of 
the observed persistence lengths difficult.

{\em \bf Model} In the present work we elucidate the meaning of $\ell_p$ for 
experimentally observed semiflexible polymers by means of Molecular Dynamics 
simulations using a bead-spring model as studied previously in both $d=2$ 
\cite{Huang} and in $d=3$ \cite{Egorov}, assuming dilute solutions under good 
solvent conditions. All beads interact with a truncated and shifted 
Lennard-Jones potential,
 \begin{figure}
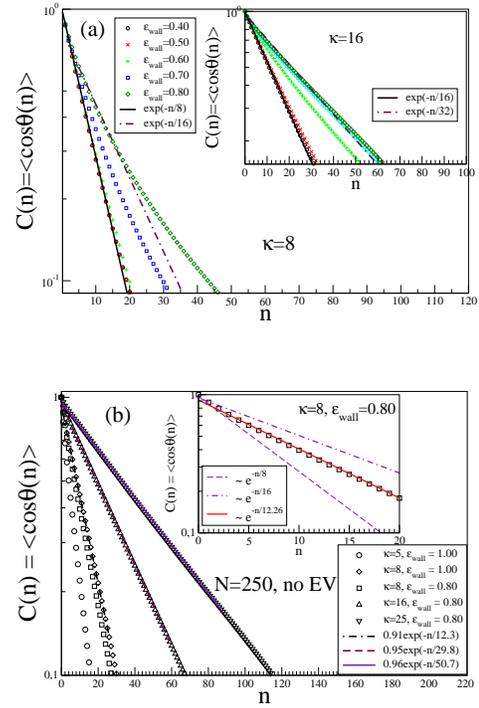
 
   \begin{adjustbox}{center}
 \includegraphics[scale=0.25, angle=0]{ACF_bond_N250_semilog.eps}
 \end{adjustbox}
 \vspace{0.5cm}
 
  \begin{adjustbox}{center}
  \includegraphics[scale=0.25, angle=0]{offset_ACF_BOND_N250_modif.eps}
  \end{adjustbox}
 \caption{(a) Semilog plot of $C(n) = \langle \cos 
\theta(n) \rangle$, vs $n$ in semi-log coordinates for $\kappa = 8$ (main 
panel) and $\kappa = 16$ (inset). Several choices of $\epsilon_{wall}$ are 
shown, as indicated. All data are for $N=250$. (b) The same as in (a) but for 
strongly adsorbed ($\epsilon_{wall}=0.80, 1.00$) chains with stiffness $\kappa = 
5, 8, 16, 25$ without EV interactions. The inset indicates the gradual 
crossover in the decay of $C(n)$ with $n$ for $\kappa = 8$ and $n=1, \; 2$ from 
$\ell_p^{eff} \approx 8$ to $\ell_p^{eff} = 12.3$ for large $n$. 
\label{fig_ACF}}
\par\vspace{-0.7cm}\par
\end{figure}

\begin{equation} \label{LJ}
 U_{LJ}(r) = 4\epsilon \left[ \left( \frac{\sigma}{r}\right)^{12} -  \left( 
\frac{\sigma}{r}\right)^{6} + \frac{1}{4}\right], 
\end{equation}
where $U_{LJ}=0$ for distances $r > 2^{1/6}\sigma$, $\epsilon$ being chosen as 
unity, $\epsilon = k_BT = 1$, and the range $\sigma = 1$.  Eq.(\ref{LJ}) 
therefore means that excluded volume effects are fully accounted for. Chain 
connectivity is ensured by the finitely extensible nonlinear elastic (FENE) 
potential \cite{Grest}, $U_{FENE}(r) = -(1/2)kR_0^2 \ln(1-r^2/R_0^2)$, with $R_0 
= 1.5 \sigma, k=30$ (the average bond length $\ell_b$ is then roughly $0.976$). 
The bond bending potential is taken as $U_b = \kappa (1-\cos\theta) \approx 
\frac{1}{2}\kappa \theta^2$, compatible with Eq.(\ref{H}), $\theta$ being the 
angle between subsequent bonds. 

A popular measure of $\ell_p$ then is \cite{Hsu5} $\ell_b/\ell_p = - \ln \langle 
\cos \theta \rangle \approx \frac{1}{2} \langle \theta^2 \rangle$, for $\kappa 
\gg 1$. This relationship yields the results displayed in 
Fig.\ref{fig_snapshot}b, i.e., $\ell_p/\ell_b \approx \kappa$, irrespective of 
the chosen substrate potential 

\begin{equation} \label{Mie}
 U_{wall}(z) = \epsilon_{wall} \left( \frac{5}{3} \right) \left( \frac{5}{2} 
\right)^{\frac{2}{3}}  \left[ \left( \frac{\sigma}{z}\right)^{10} -  
\left( \frac{\sigma}{z}\right)^{4}  \right],
\end{equation}
which has a minimum $U_{wall}(z_{min}) = - \epsilon_{wall}$ at $z_{min} / \sigma 
= \left(\frac{5}{2}\right)^{1/6}$. In the simulations below, varying 
$\epsilon_{wall}$ and the chain length $N$, we have carefully monitored that on 
the available time scale (of the order of up to $10$ million MD time units) 
equilibrium is reached. In each case $50$ runs (carried out in parallel using 
graphics processing unit) were averaged over.

{\bf Results} While for small $\epsilon_{wall}$ the chains are essentially 
non-adsorbed mushrooms (one chain end being fixed at the surface), for 
$\epsilon_{wall} \approx 1.0$ all monomers are bound to the wall, i.e., a 
quasi-twodimensional conformation occurs. Surprisingly, for neighboring bonds, 
$s-s'=1$, the expected change of the effective decay length $\ell_p^{eff}$ of 
orientational correlations from $\ell_p$ to $2\ell_p$ ($2\ell_p$ is readily seen 
for strictly $d=2$ chains \cite{Huang}) is {\it not} observed.

This finding is rationalized by considering two subsequent bonds, the first bond 
from $\vec{r}_{j-1}$ to  $\vec{r}_{j}$, the second from $\vec{r}_{j}$ to 
$\vec{r}_{j+1}$, (cf. inset to Fig.\ref{fig_snapshot}b). Choosing polar 
coordinates to describe the bonds $\vec{r}_{j} - \vec{r}_{j-1} = \ell_b (-\cos 
\alpha_j, 0, \sin \alpha_j)$ and $\vec{r}_{j+1} - \vec{r}_{j} = \ell_b 
(\cos\alpha_{j+1} \cos\phi, \cos\alpha_{j+1} \sin\phi,\sin\alpha_{j+1})$, for 
small angles $\theta$ between the bonds one has $\theta^2 = \phi^2 + (\alpha_j - 
\alpha_{j+1})^2$, therefore, also for an adsorbed polymer the bond angle 
$\theta$ is composed from {\it two} transverse degrees of freedom. Only if the 
wall potential would constrain all positions $\{z_j\}$ strictly to $z_{min}$ , 
one would get $\alpha_j - \alpha_{j+1} \equiv 0$, that is, a {\it single} 
transverse degree of freedom. There are slight deviations from the result 
$\ell_p/\ell_b =\kappa$ in Fig.\ref{fig_snapshot}b. However, when one follows 
$\langle \cos \theta(n)\rangle$ for large distances $n$ along the contour,  
Fig.\ref{fig_ACF}a, one reproduces Eq.(\ref{lp}) strictly only for the 
non-adsorbed mushrooms, for all the weakly adsorbed chains, instead, the strong 
curvature of the semilog plot shows that an interpretation by Eq.(\ref{lp}) with 
a single decay length is inadequate. While quantitative details in 
Figs.\ref{fig_snapshot}, \ref{fig_ACF} depend on the specific chain model and 
the wall potential, the fact that $\langle \cos \theta(n) \rangle$ is not 
compatible with Eq.(\ref{lp}) for weakly adsorbed chains even at large $n$, and 
for strongly adsorbed chains applies only if both $\ell_p$ and $n$ are large, is 
a generic feature.  For strongly adsorbed chains a crossover of the effective 
decay length $\ell_p$ to about $2\ell_p$ occurs when $n$ is significantly larger 
than $1$. The further crossover to the power law \cite{Hsu3} $C(n) \propto 
n^{-1/2}$ for $n \gg 2\ell_p$ in Fig.~\ref{fig_ACF}a sets in slowly, the fully 
developed power law is not seen here, it would require to study by far longer 
chains.  To separate the EV effect from the crossover $\ell_p \to 2\ell_p$ 
caused by adsorption, we simulate chains where Eq.(\ref{LJ}) between non-bonded 
monomers was omitted, Fig.\ref{fig_ACF}b. One sees that $\ell^{eff}_p$  reaches 
the value $2\ell_p$ only for large $\kappa$.

\begin{figure}[htb]
 \vspace{0.5cm}
 \centering
  \begin{adjustbox}{center}
 \includegraphics[scale=0.27, angle=0]{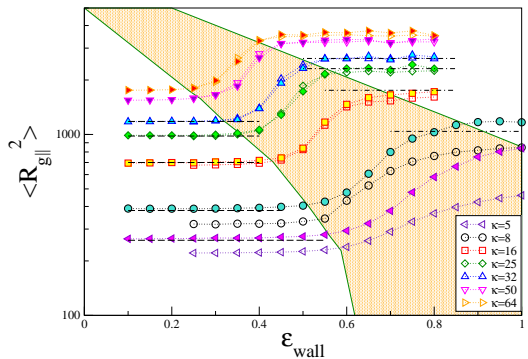}
\end{adjustbox}

\caption{Mean-square lateral gyration radius for  $N=250$ vs $\epsilon_{wall}$ 
for $7$ choices of $\kappa$ with (full symbols) and without (open symbols) 
excluded volume (EV) interactions. EV is more important for small $\kappa$ 
($\kappa = 5$, $8$). Horizontal straight lines show KP predictions for $d=3$ 
(for small $\epsilon_{wall}$) and $d=2$ (for larger $\epsilon_{wall}$). The 
shaded transition region from non-adsorbed to adsorbed chains narrows down with 
growing $\kappa$. \label{fig_Rgxy}}
\par\vspace{-0.7cm}\par
 \vspace{0.5cm}
\end{figure}

\begin{figure}[h]
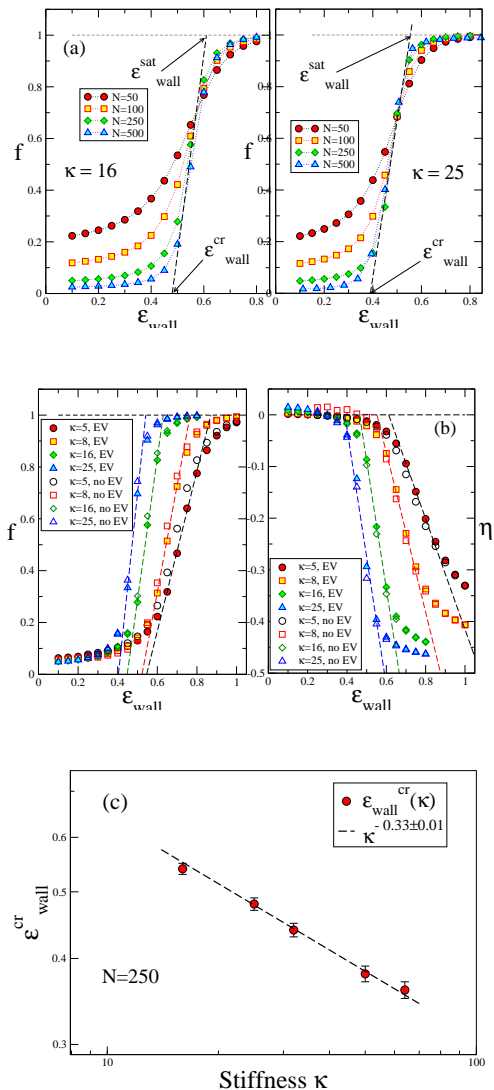

\centering
 \begin{adjustbox}{center}
 \includegraphics[scale=0.25, angle=0]{Fig.3a.eps}
   \end{adjustbox}
   
 \vspace{0.7cm}
  \begin{adjustbox}{center}
 \includegraphics[scale=0.25, angle=0]{Fig.3b.eps}
   \end{adjustbox}
   
 \vspace{0.7cm}
  \begin{adjustbox}{center}
 \includegraphics[scale=0.25, angle=0]{CAP_k_N250.eps}
 \end{adjustbox}
 \caption{(a) Fraction $f$ of adsorbed monomers vs $\epsilon_{wall}$ for $\kappa 
= 16$ (left panel) and $\kappa = 25$ (right panel), for the $4$ chain lengths 
$N=50, 100, 250$ and $500$, respectively. Tentative linear extrapolations 
indicate the estimated location of the adsorption transition, 
$\epsilon^{cr}_{wall}$, (nonzero $f$ for $\epsilon_{wall} < 
\epsilon^{cr}_{wall}$ is a finite-size effect). Also the estimation of 
$\epsilon^{sat}_{wall}$, where $f$ crosses over to saturation value $f=1$, is 
indicated. (b) Adsorbed fraction $f$ plotted vs $\epsilon_{wall}$ for $N=250$ 
and $4$ choices of $\kappa$ (left), and orientational order parameter of the 
bonds $\eta = \frac{3}{2}\langle \cos^2 \vartheta \rangle - \frac{1}{2}$ vs 
$\epsilon_{wall}$ (right). Data with no EV interaction (open symbols) are also 
included. (c) Variation of the critical adsorption potential 
$\epsilon_{wall}^{cr}$ with chain stiffness $\kappa$ for $N=250$. 
\label{fig_OP}}
\par\vspace{-0.7cm}\par
\end{figure}

The gradual crossover from $e^{-n/\ell_p}$ to $e^{-n/2\ell_p}$ with increasing 
$\epsilon_{wall}$, and the precise range of $\epsilon_{wall}$ where this occurs, 
reflect the region over which the adsorption transition is rounded (owing to the 
finite chain length $N$) and depend on $\kappa$ as well. The rounded transition 
is monitored by studying the lateral chain linear dimensions, 
Fig.\ref{fig_Rgxy}, or local order parameters, the fraction $f$ of adsorbed 
monomers, defined by $f = \int\rho(z)U_{wall}(z)dz / \int U_{wall}(z)dz$, 
Fig.\ref{fig_OP}a, or the orientational order parameter of the bonds $\eta = 
\frac{3}{2}\langle \cos^2\vartheta \rangle - \frac{1}{2}$, $\vartheta$ being the 
angle of a bond with the surface normal,  Fig.\ref{fig_OP}b. For the shown 
medium chain lengths, EV effects for non-adsorbed chains are negligible for all 
$\kappa$. They are, however, present for $N=100$ for adsorbed chains with 
$\kappa=5$ and $8$ whereas for $N=250$ also data for adsorbed chains with 
$\kappa = 16$ and $\kappa = 25$ are already slightly affected by excluded 
volume. These findings are certainly compatible with experiment: for ss-DNA with 
$\ell_p \approx  4.6$ to $9.1 nm$, depending on the ion concentration in the 
solution, evidence for $\langle R^2(s) \rangle \propto s^{2\nu}$ with $\nu = 
0.73$ was presented \cite{Rechendorff}, in contrast to the KP  prediction  
$\langle R^2(s) \rangle \propto \ell_p s$. Even for long enough ds-DNA with 
$\ell_p = 50 nm$ (with effective diameter $\sigma = 2 nm$, this would correspond 
to $\kappa = 25$ in our model), the $d=2$  SAW-type behavior was observed 
clearly \cite{Radler}. Thus, the suggestion \cite{Moukhtar1} to estimate 
$\ell_p$ from the KP expression by means of AFM measurement on DNA in the limit 
$L \gg \ell_p$ must be taken with due care  since significant systematic errors 
may occur when both $L$ and $\ell_p$  are used as adjustable parameters. 

Fig.\ref{fig_Rgxy} also includes a rough estimate of the lateral part of the 
mean-square gyration radius of non-adsorbed mushrooms ($\langle R^2_{gxy} 
\rangle \equiv \langle R^2_{gx}\rangle + \langle R^2_{gy}\rangle \approx 2/3 
\langle R^2_g \rangle^{KP}_{d=3}$ (with $\langle R^2_g \rangle^{KP}_{d=3}$ being 
the result of the KP model in $d=3$). For large $\epsilon_{wall}$, the data 
roughly converge towards the corresponding predictions in $d=2$ dimensions 
$\langle R^2_g \rangle^{KP}_{d=2}$ (provided $\kappa$ is large enough too). 
Denoting $n_p \equiv N/\ell_p$ (where $\ell_p$ is the $d=3$ persistence length), 
one has in $d=2$:
\begin{equation} \label{Rg_KP}
 \frac{3\langle R^2_g\rangle}{2\ell_p L} = 1 - \frac{6}{n_p} \left \{1 -  
\frac{4}{n_p} \left [1 - \frac{4}{n_p} \left(1- \frac{1-\exp(-n_p/2)}{n_p/2} 
\right)\right]\right\},
\end{equation}
whereas in $d=3$ the same expression holds yet with  $\ell_p$ being replaced by 
$\ell_p/2$ (also in $n_p$).

It is clear from Fig.\ref{fig_Rgxy} that the KP model, Eq.(\ref{Rg_KP}), is 
inapplicable in the broad (shaded) transition region from weakly to strongly 
adsorbed chains. The chain conformations contain here large loops (whereby 
$\ell_p$ appropriate for $d=3$ applies) as well as some trains (where 
$\ell_p^{eff} \approx 2\ell_p$). But even if the chains are so strongly adsorbed 
that loops no longer occur, $\ell^{eff}_p$  is less than $2\ell_p$ for 
intermediate values of $\kappa$, as Fig.\ref{fig_ACF}b shows: A decay law $C(n) 
= A \exp(-n/ \ell_p^{eff})$ is observed, with $A < 1$ and $\ell_p^{eff} < 
2\ell_p$. Using these results to modify Eq.(\ref{Rg_KP}), we can account for the 
actual values of $\langle R_g^2 \rangle$ in the strongly adsorbed regime shown 
in Fig.\ref{fig_Rgxy} for those chains where EV is switched off. Thus, e.g., for 
$\kappa = 8, \epsilon_{wall} = 1, N = 250$, Eq.(\ref{Rg_KP}) would yield a 
$\langle R_g^2 \rangle$ of $1039$ while the observation is about $847$ only. 
Taking into account that $n_p^{eff} = 36.55$ instead of $n_p =31.125$, and the 
reduction by $A = 0.93$ (see Fig.\ref{fig_ACF}b), we predict $845$, in very good 
agreement with the simulation. Of course, for such not very stiff and rather 
long chains the complete neglect of excluded volume is not warranted, as $R_g^2 
= 1167$ for the chain with EV shows. As seen in Fig.\ref{fig_ACF}a, EV also 
causes onset of curvature in the semilog plot of $C(n)$. Thus Gaussian 
statistics as implicit in the KP model for $L \gg l_p$ is clearly  inadequate. 
Only for all the data without EV, the modified KP model (with $\ell_p^{eff}$ 
rather than $2\ell_p$) can account for the results qualitatively.

Since the adsorption transition becomes a well-defined (sharp) phase transition 
only for $N \to \infty$, and then the theory predicts\cite{Semenov} that $f 
\propto (\epsilon_{wall} - \epsilon_{wall}^{cr})$ for $\epsilon_{wall}^{cr} < 
\epsilon_{wall} < \epsilon_{wall}^{sat}$ for semiflexible polymers, we plot $f$ 
vs $\epsilon_{wall}$ in Fig.\ref{fig_OP}a for $N$ extending from $N=50$ to 
$N=500$. Indeed, the data are qualitatively compatible with this prediction, and 
the estimates, $\epsilon_{wall}^{cr}$, thus obtained comply within error bars 
with the predicted\cite{Semenov} behavior $\epsilon_{wall}^{cr}/k_BT \propto 
(\ell_p/\ell_b)^{-1/3} = \kappa^{-1/3}$, cf. Fig.\ref{fig_OP}c. This is 
understood qualitatively by decomposing the adsorbed chain into straight pieces 
of length $\lambda \propto \ell_p^{1/3} \Delta^{2/3}$, $\Delta$ being the range 
of the adsorption potential while $\lambda$ is the ``deflection length`` 
\cite{Odijk}. The transition occurs when the energy won by one such piece is of 
order $k_BT$. Fig.\ref{fig_OP}a also shows estimates of $\epsilon^{sat}$ where 
$f$ gradually reaches saturation, $f \to 1$. However, while with increasing 
$\kappa$ the curves $f$ vs $\epsilon_{wall}$ do become steeper, we are still far 
from the $1^{st}$-order-like behavior, predicted \cite{Semenov} for $\kappa \to 
\infty$.

{\em \bf Conclusions}
In summary, using a bead-spring model with a bond-angle potential where the 
nonbonded part of the excluded volume potential between monomers is either 
included or switched off, a test of the KP description of the adsorption of 
semiflexible polymers is presented. Unlike previous lattice model work 
(predicting $\epsilon_{wall}^{cr} \propto 1/\ell_p$), we verify Semenov's [17] 
prediction $\epsilon_{wall}^{cr} \propto 1/\ell_p^{1/3}$. Ref.\cite{Semenov} 
presents a precise description of the adsorption of ideal wormlike (KP) chains 
and explains why previous attempts (apart from considerations based on the 
unbinding transitions \cite{Maggs}) failed.  While near the transition (for very 
stiff chains) excluded volume is unimportant, it matters for strongly adsorbed 
quasi-$2d$ chains. We show that the concept of persistence length is not useful 
for weakly adsorbed chains, and for the strongly adsorbed chains we demonstrate 
that the $\ell_p \to 2\ell_p$ change, predicted by the KP model, only holds for 
very large $\ell_p$. We expect that these findings will help the proper 
interpretation of experiments on adsorbed ss-DNA and ds-DNA.

{\em \bf Acknowledgements}
A.M. is indebted to the Alexander von Humboldt foundation for financial support 
during this study and also thanks the COST action No. CA17139, supported by COST 
(European Cooperation in Science and Technology \cite{COST}) and its Bulgarian 
partner FNI/MON under KOST-11.


\begin{thebibliography}{10}

\bibitem{Wiggins} P. A. Wiggins, T. van der Heijden, F. Moreno-Herrero, A. 
Spakowitz, R. Phillips, J. Widom, C. Dekker, and P. C. Nelson, Nature 
Nanotechnology {\bf 1}, 137 (2006)
\bibitem{Han} L. Han, H. G. Garcia, S. Blumberg, K. B. Towles, J. F. Beausang, 
P. C. Nelson, R. Phillips, PLOS One {\bf 4}, e5621 (2009)
\bibitem{Ha} E. Vafabaksch and T. Ha, Science {\bf 337}, 1097 (2012)
\bibitem{Moukhtar1} J. Moukhtar, E. Fontaine, C. Faivre-Moskalenko, and A. 
Arneodo, Phys. Rev. Lett. {\bf 98}, 178101 (2007)
\bibitem{Moukhtar2} J. Moukhtar, C. Faivre-Moskalenko, P. Milani, B. Audit, 
C. Vaillant, E. Fontaine, F. Mongelar, G. Lavorel, P. St-Jean, P. Bouvet, 
F. Argoul, and Alain Arneodo, J. Phys. Chem. B {\bf 114}, 5125 (2010)
\bibitem{Radler} B. Maier and J. O. R\"{a}dler, Phys. Rev. Lett. {\bf 82}, 1911 
(1999)
\bibitem{Rechendorff} K. Rechendorff, G. Witz, J. Adamcik, and G. Dietler, J. 
Chem. Phys. {\bf 131}, 095103 (2009) 
\bibitem{Sinha} R. Sinha,H. L. Frisch, and F. R. Eirich,  J. Chem. Phys. {\bf 
57}, 584 (1953)
\bibitem{deGennes} P. G. de Gennes, Macromolecules {\bf 14}, 1637 (1981)
\bibitem{Eisenriegler} E. Eisenriegler,  K. Kremer, and K. Binder, J. Chem. 
Phys. {\bf 77}, 6296 (1982)
\bibitem{Fleer} G. J. Fleer,  M. A. Cohen Stuart, J. M. H. M. Scheutjens, T. 
Cosgrove, and B. Vincent, {\it Polymers at Interfaces} (Chapman and Hall, 
London, 1993)
\bibitem{Klushin} L. I. Klushin, A. A. Polotsky, H.-P. Hsu, D. A. Markelov, K. 
Binder, and A. M. Skvortsov,  Phys. Rev. E {\bf 87}, 022604 (2013)
\bibitem{Birshtein} T. M. Birshtein,  E. B. Zhulina, A. M. Skvortsov, 
Biopolymers {\bf 18}, 1171 (1979)
\bibitem{Khokhlov} A. R. Khokhlov, F. F. Ternovsky, and E. A. Zheligovskaya, 
 Macromol. Chem. Theory Simul. {\bf 2}, 151 (1993)
\bibitem{Kramarenko} E. Yu. Kramarenko, R. G. Winkler, P. G. Khalatur,  A. R. 
Khokhlov, and P. Reineker, J. Chem. Phys. {\bf 104}, 4806 (1996)
\bibitem{Maggs} A. C. Maggs, D. A. Huse, and S. Leibler, Europhys. Lett. {\bf 
8}, 615 (1989)
\bibitem{Semenov} A. N. Semenov, Eur. Phys. J. E {\bf 9}, 353 (2002)
\bibitem{Netz} R. R. Netz and D. Andelman, Phys. rep. {\bf 380}, 1 (2003)
\bibitem{Deng} M. Deng, Y. Jiang, H. Liang, and J. Z. Y. Chen, J. Chem. Phys. 
{\bf 133}, 034902 (2010)
\bibitem{Hsu1} H.-P. Hsu and K. Binder, Macromolecules, {\bf 46}, 2496 (2013)
\bibitem{Kampmann} T. A. Kampmann, H. H Boltz, and J Kierfeld, J. Chem. Phys. 
{\bf 139}, 034903 (2013) 
\bibitem{Baschnagel} J. Baschnagel, H. Meyer, J. Wittmer, I. Kuli\'{c}, 
H. Mohrbach, F. Ziebert, G-M. Nam, N.-K. Lee, and A. Johner, Polymers {\bf 8}, 
286 (2016)
\bibitem{Kierfeld} T. A. Kampmann and J. Kierfeld, J. Chem. Phys. {\bf 147}, 
014901 (2017)
\bibitem{Grosberg} A. Grosberg and A.R. Khokhlov, {\it Statistical Physics of 
Macromolecules} (AIP Press, New York, 1994)
\bibitem{Odijk} T. Odijk, Macromolecules, {\bf 16}, 1340 (1983)
\bibitem{KP} O. Kratky and G. Porod, J. Colloid Sci., {\bf 4} 35 (1949) 
\bibitem{Hsu2} H.-P. Hsu, W. Paul, and K. Binder, EPL {\bf 92}, 28003 (2010)
\bibitem{Hsu3} H.-P. Hsu, W. Paul, and K. Binder,  EPL {\bf 95}, 68004 (2011)
\bibitem{Lee} N.-K. Lee and A. Johner, Macromolecules {\bf 48}, 7681 (2015)
\bibitem{Grebikova} L. Grebikova, S. Kozhuharov, P. Maroni, A. Mikhaylov, G. 
Dietler,  A. D. Schl\"{u}ter, M. Ullnerd,  and  M. Borkovec, Nanoscale {\bf 8}, 
13498 (2016)
\bibitem{Hsu4} H.-P. Hsu, W. Paul, and K. Binder, J. Chem. Phys. {\bf 133}, 
134902 (2010) 
\bibitem{Hsu5} H.-P. Hsu, W. Paul, and K. Binder,  Macromolecules {\bf 43}, 
3094 (2010)
\bibitem{Messner} D. Messner, Christoph B\"{o}ttcher, H. Yu, A. Halperin, 
Kurt Binder, Martin Kr\"{o}ger, and A. D. Schl\"{u}ter,  ACS~NANO, {\bf 13}, 
 3466 (2019)
\bibitem{Duterte} F. Dutertre, Ki-Taek Bang, E. Vereroudakis, B. Loppinet, 
Sanghee Yang, Sung-Yun Kang, George Fytas, and Tae-Lim Choi, Macromolecules {\bf 
52}, 3342 (2019) 
\bibitem{Huang} A. Huang, H.-P. Hsu, Aniket Bhattacharya, and K. Binder, J. 
Chem. Phys. {\bf 143}, 243102 (2015)
\bibitem{Egorov} S. A. Egorov, A. Milchev, P. Virnau, and K. Binder, Soft Matter 
{\bf 12}, 4944 (2016)
\bibitem{Grest} G. S. Grest and K. Kremer, Phys. Rev. A {\bf 33}, 3628 (1986)
\bibitem{COST} See http://www.cost.eu and https://www.fni.bg 
\end{thebibliography}
\end{document}